\documentclass[aps,singlecolumn]{revtex4-1}
\usepackage{graphicx}
\usepackage{times}
\usepackage{amssymb,amsfonts,amsmath}
\usepackage{color}
\newcommand{\beq}{\begin{equation}}
\newcommand{\eeq}{\end{equation}}
\newcommand{\bea}{\begin{eqnarray}}
\newcommand{\eea}{\end{eqnarray}}

\newcommand{\affilb}{
 ELTE-MTA ``Lendulet'' Biophysics Research Group,
 Department of Biological Physics, E\"otv\"os University,
 P\'azm\'any P.\ stny.\ 1A, H-1117 Budapest, Hungary
}

\newcommand{\affila}{
 ELTE-MTA ``Lendulet'' Evolutionary Genomics Research Group,
 Department of Biological Physics, E\"otv\"os University,
 P\'azm\'any P.\ stny.\ 1A, H-1117 Budapest, Hungary
}

\begin{document}

\title{Hierarchical tissue organization as a general mechanism to limit
the accumulation of somatic mutations}

\author{Imre Der\'enyi}
\email[]{derenyi@elte.hu}
\affiliation{\affilb}
\author{Gergely J. Sz\"oll\H{o}si}
\email[]{ssolo@elte.hu}
\affiliation{\affila}

\begin{abstract}
\begin{center}
\large \bf Abstract
\end{center}

How can tissues generate large numbers of cells, yet keep
the divisional load (the number of divisions along
cell lineages) low in order to curtail the accumulation of somatic
mutations and reduce the risk of cancer?
To answer the question we consider a general model of hierarchically organized self-renewing tissues and show that the lifetime divisional load of such a tissue is independent of the details of the cell differentiation processes, and depends only on two structural and two dynamical parameters.
Our results demonstrate that a strict analytical relationship exists between two seemingly disparate characteristics of self-renewing tissues: divisional load and tissue organization.
Most remarkably, we find that a sufficient number of progressively slower dividing cell types can be almost as efficient in minimizing the divisional load, as non-renewing tissues.
We argue that one of the main functions of tissue-specific stem cells and differentiation hierarchies is the prevention of cancer.
\end{abstract}
\maketitle

\section*{Introduction}
\begin{figure}
\centerline{\includegraphics[width=0.6\textwidth]{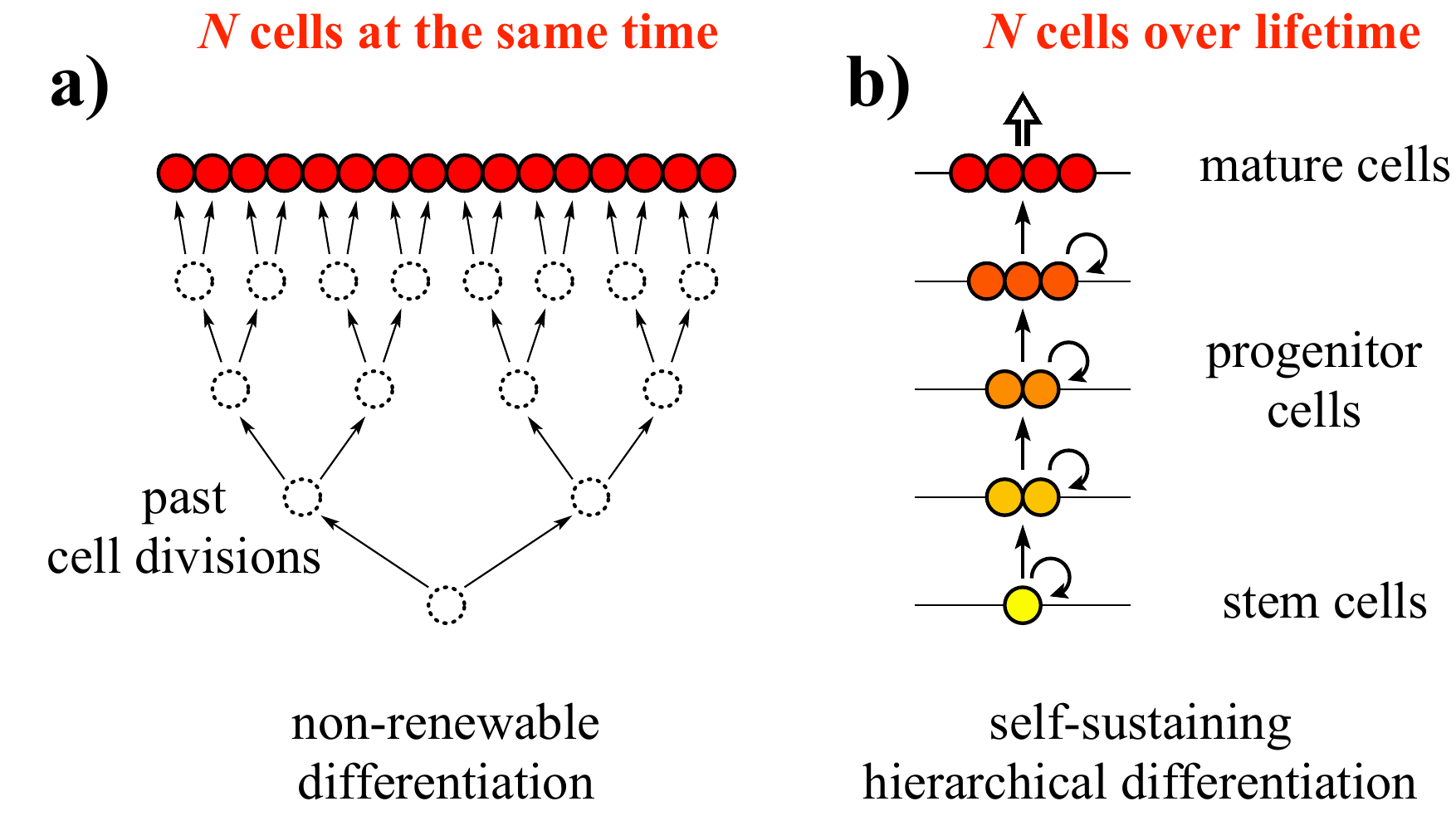}}
\caption{
{\bf Differentiation in non-renewing vs.\ self-renewing tissues.}
a) To produce $N$ mature cells from a single precursor with a minimum
number of cell divisions, $\log_2(N)$, strict division along a perfect
binary tree is necessary. In multicellular organisms such ``non-renewable''
differentiation typically takes place early in development. b) However, in
self-renewing tissues, where homeostasis requires a continuous supply
of cells, a small population of self-replicating tissue-specific stem
cells sustain a hierarchy of progressively differentiated and larger
populations of cell types, with cells of each type being continuously
present in the tissue.
}
\label{fig1}
\end{figure}

In each multicellular organism a single cell proliferates to produce
and maintain tissues comprised of large populations of differentiated
cell types. The number of cell divisions in the lineage leading to a
given somatic cell governs the pace at which mutations accumulate
\cite{Gao:2016}. The resulting somatic mutational load determines the
rate at which unwanted evolutionary processes, such as cancer
development, proceed \cite{Nowell:1976,Merlo:2006,Beerenwinkel:2016}.
In order to produce $N$ differentiated cells from a single precursor
cell the theoretical minimum number of cell divisions required along
the longest
lineage is $\log_2(N)$. To achieve this theoretical minimum, cells must
divide strictly along a perfect binary tree of height $\log_2(N)$
(Fig.~\ref{fig1}a). In
multicellular organisms such differentiation typically
takes place early in development. It is responsible for producing the
cells of non-renewing tissues (e.g., primary oocytes in the female germ
line \cite{Crow:2000,Gao:2016}) and the initial population of stem
cells in self-renewing tissues (e.g., hematopoietic stem cells
\cite{Busch:2015,Werner:2015,Werner:2015_eLife}
or the spermatogonia of the male germ line
\cite{Crow:2000,Gao:2016}).

In self-renewing tissues, which require a continuous supply of cells,
divisions along a perfect binary tree are unfeasible. Strictly following
a perfect binary tree throughout the lifetime of the organism would require
extraordinarily elaborate scheduling of individual cell divisions to
ensure tissue homeostasis \cite{Morris:2014}, and would be singularly
prone to errors (e.g., the loss of any single cell would lead to the
loss of an entire branch of the binary tree). Instead, to compensate
for the continuous loss of cells, mechanisms have evolved to replenish
the cell pool throughout the organism's lifetime \cite{Pardee:1989}. In
most multicellular organisms hierarchically organized tissue structures
are utilized. At the root of the hierarchy are a few tissue-specific
stem cells defined by two properties: self-replication and the
potential for differentiation \cite{Till:1961,McCulloch:2005}. During
cell proliferation cells can differentiate and become increasingly
specialized toward performing specific functions within the hierarchy,
while at the same time losing their stem cell-like properties
(Fig.~\ref{fig1}b). A classic example is the hematopoietic
system \cite{Michor:2005,Dingli:2007}, but other tissues such as skin
\cite{Tumbar:2004} or colon \cite{Barker:2007,Potten:2009} are also
known to be hierarchically organized. Identifying each level of the
hierarchy, however, can be difficult, especially if the cells at
different levels are only distinguished by their environment, such as
their position in the tissue (e.g., the location of the
transit-amplifying cells along intestinal crypts). As a result,
information on the details of differentiation hierarchies is incomplete
\cite{Rossi:2008,Vermeulen:2013,Sutherland:2015}.

Nonetheless, in a recent paper, Tomasetti and Vogelstein
\cite{Tomasetti:2015} gathered available information from the
literature and investigated the determinants of cancer risk among
tumors of different tissues. Examining cancers of 31 different tissues
they found that the lifetime risk of cancers of different types is
strongly correlated with the total number of divisions of the normal
self-replicating cells. Their conclusion that the majority of cancer
risk is attributable to bad luck \cite{Tomasetti:2015} arguably results
from a misinterpretation of the correlation between the logarithms of
two quantities \cite{Wild:2015,Wu:2016}. However, regardless of the
interpretation of the correlation, the data display a striking
tendency: the dependence of cancer incidence on the number of stem cell
divisions is sub-linear, i.e., a 100 fold increase in the number of
divisions only results in a 10 fold increase in incidence. This
indicates that tissues with a larger number of stem cell divisions
(typically larger ones with rapid turnover, e.g., the colon) are
relatively less prone to develop cancer. This is analogous to the
roughly constant cancer incidence across animals with vastly different
sizes and life-spans (Peto's paradox), which implies that large animals
(e.g., elephants) possess mechanisms to mitigate their risk relative to
smaller ones (e.g., mice) \cite{Peto:1975,Caulin:2011,Peto:2015}.

What are the tissue-specific mechanisms that explain the differential
propensity to develop cancer?
It is clear that stem cells that sustain hierarchies of progressively
differentiated cells are well positioned to provide a safe
harbor for genomic information.
Qualitative arguments suggesting that hierarchically organized
tissues may be optimal in reducing the accumulation of somatic
mutations go back several decades \cite{Hindersin:2016}.
As mutations provide the fuel for somatic evolution (including not only
the development of cancer, but also tissue degeneration, aging, germ
line deterioration, etc.) it is becoming widely
accepted that tissues have evolved to minimize the accumulation of
somatic mutations during the lifetime of an individual
\cite{Hindersin:2016}.
The potential of hierarchical
tissues to limit somatic mutational load simply by reducing the number of cell
divisions along cell lineages, however, has not been explored in a
mathematically rigorous way. Here, we discuss this most fundamental mechanism by which hierarchical
tissue organization can curtail the accumulation of somatic
mutations. We derive simple and general analytical properties of the divisional load of a
tissue, which is defined as the number of divisions its constituent
cells have undergone along the longest cell lineages,
and is expected to be proportional to the
mutational load of the tissue.


Models conceptually similar to ours have a long history
\cite{Loeffler:1980,Nowak:2003,Takizawa:2011sf,Pepper:2007,Werner:2011,Werner:2013,Werner:2015},
going back to Loeffler and Wichman's work on modeling
hematopoietic stem cell proliferation \cite{Loeffler:1980},
and several qualitative arguments have been made suggesting why
hierarchically organized tissues may be optimal in minimizing somatic
evolution. In a seminal contribution Nowak et al.\ \cite{Nowak:2003}
showed that tissue architecture
can contribute to the protection against the accumulation of somatic
mutations. They demonstrated that the rate of somatic evolution will be
reduced in any tissue where geometric arrangement or cellular
differentiation induce structural asymmetries such that mutations that
do not occur in stem cells tend to be washed out of the cell
population, slowing down the rate of fixation of mutations.
Here, we begin where Nowak et al. \cite{Nowak:2003} left off: aside of
structural asymmetry, we consider a
second and equally important aspect of differentiation, the dynamical
asymmetry of tissues, i.e., the uneven distribution of divisional rates
across the differentiation hierarchy.

More recently a series of studies have investigated the
dynamics of mutations in hierarchical tissues with dynamical
asymmetry \cite{Pepper:2007,Werner:2011,Werner:2013}
and found that hierarchical tissue organization can
(i) suppress single \cite{Werner:2011} as well as multiple mutations
\cite{Werner:2013} that arise in progenitor cells, and
(ii) slow down the rate of somatic evolution towards cancer
\cite{Pepper:2007} if selection on mutations with
non-neutral phenotypic effects is also taken into account.
The epistatic interactions between individual driver mutations are,
however, often unclear and show large variation among cancer types.
The fact that the majority of cancers arise without a histologically
discernible premalignant phase indicates strong cooperation between
driver mutations, suggesting that major histological changes may not
take place until the full repertoire of mutations is acquired
\cite{Martincorena:2015rev}. For this reason, here we do not consider
selection between cells, but rather, focus only on the pace of the
accumulation of somatic mutations in tissues, which provide the fuel
for somatic evolution.

The uneven distribution of divisional rates considered by Werner et al.\
\cite{Werner:2011,Werner:2013}
followed a power law, however, this distribution was taken
for granted without prior justification. Their focus was instead on
``reproductive capacity'', an attribute of a single cell corresponding
to the number of its descendants, which is conceptually unrelated to
our newly introduced ``divisional load'', which characterizes the
number of cell divisions along the longest cell lineages of the tissue.
Here we show mathematically, to the best of our knowledge for the first
time, that the minimization of the divisional load in hierarchical
differentiation indeed leads to power law distributed differentiation
rates.


More generally, evolutionary thinking is becoming an indispensable
tool to understand cancer, and even to propose directions in the
search for treatment strategies \cite{Komarova:2015}.
Models that integrate information on
tissue organization have not only provided novel insight into cancer
as an evolutionary process \cite{Rejniak:2011,Altrock:2015,Hindersin:2016}, but have
also produced direct predictions for improved
treatment \cite{Michor:2015,Tang:2016,Werner:2016}. The simple and
intuitive relations that we derive below have the potential to further
this field of research by providing quantitative grounds for the deep
connection between organization principles of tissues and disease
prevention and treatment.


According to our results, the
lifetime divisional load of a hierarchically organized tissue is
independent of the details of the cell differentiation processes.
We show that in self-renewing tissues hierarchical organization
provides a robust and nearly ideal mechanism to limit the divisional
load of tissues and, as a result, minimize the accumulation of somatic
mutations that fuel somatic evolution and can lead to cancer.
We argue that
hierarchies are how the tissues of multicellular organisms keep the
accumulation of mutations in check, and that populations of cells
currently believed to correspond to tissue-specific stem cells
may in general constitute a diverse set of slower dividing cell
types \cite{Li:2010,Busch:2015}. 
Most importantly, we find that the theoretical minimum number
of cell divisions can be very closely approached: as long as a
sufficient number of progressively slower dividing cell types
towards the root of the hierarchy are present,
optimal self-sustaining differentiation hierarchies can
produce $N$ terminally differentiated cells during the course of an
organism's lifetime from a single precursor with no more than
$\log_2(N)+2$ cell divisions along any lineage.

\section*{Results}

\subsection*{Divisional load of cell differentiation hierarchies}

\begin{figure}
\centerline{\includegraphics[width=0.8\columnwidth]{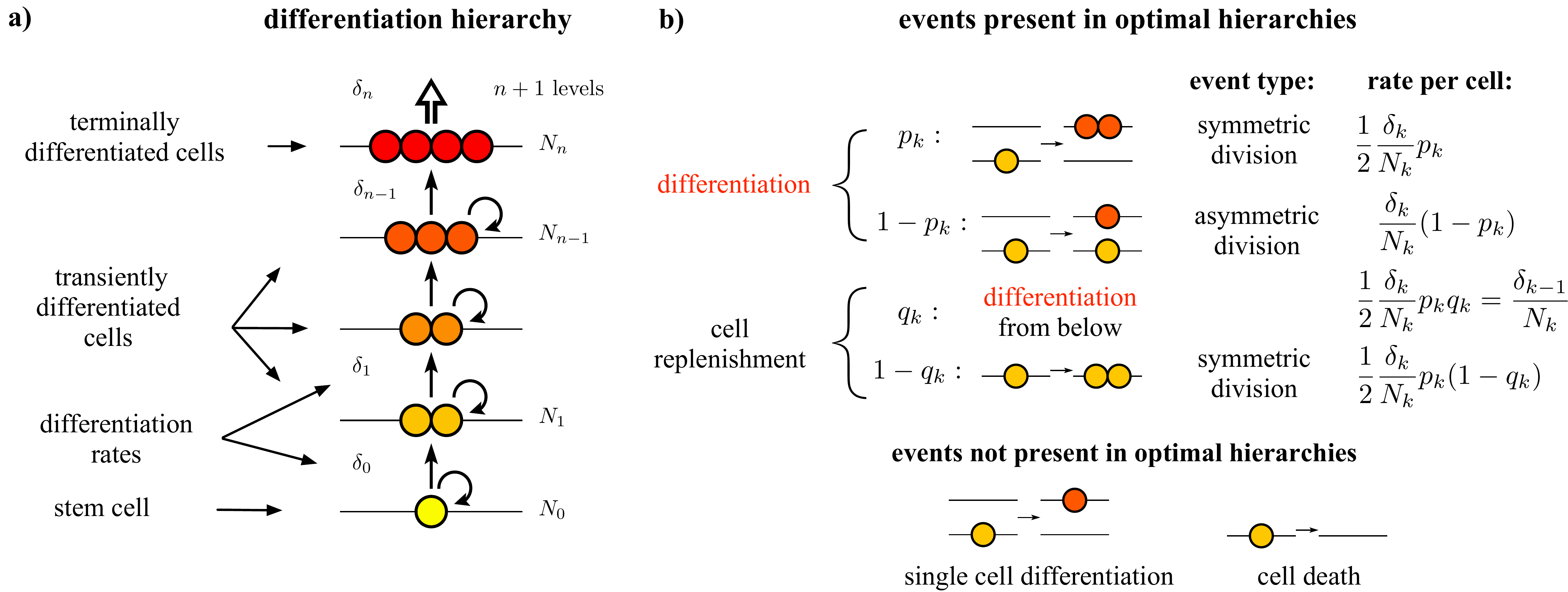}}
\caption{
{\bf Hierarchical cell differentiation in self-renewing tissue.}
a) A model tissue produces terminally differentiated cells through $n$
intermediate levels of partially differentiated cells.
b) Five microscopic events can occur with a cell:
(i) symmetric cell division with differentiation, (ii) asymmetric cell
division, (iii) symmetric cell division without differentiation, (iv)
single cell differentiation, and (v) cell death.
To the
right of each type of event present in optimal hierarchies we give the corresponding per cell rate
that is used to derive Eq.~\ref{dotD}.
}
\label{fig2}
\end{figure}

To quantify how many times the cells of self-renewing
tissues undergo cell divisions during tissue development and maintenance,
we consider a minimal generic model of hierarchically organized,
self-sustaining tissue. According to the model, cells are organized into $n+1$
hierarchical levels based on their differentiation state.
The bottom level (level $0$) corresponds to tissue-specific stem cells,
higher levels represent
progressively differentiated progenitor cells, and the top level
(level $n$) is comprised of terminally differentiated cells (Fig.~\ref{fig2}a).
The number of cells at level $k$ in fully
developed tissue under normal homeostatic conditions is denoted by $N_k$.
During homeostasis cells at
levels $k<n$ can differentiate (i.e., produce cells for
level $k+1$) at a rate $\delta_k$, and have the potential for
self-replication. At the topmost $k=n$ level of the hierarchy terminally
differentiated cells can no longer divide and are expended at the same
rate $\delta_{n-1}$ that they are produced from the level below.
The differentiation rates
$\delta_k$ are defined as the total number of differentiated
cells produced by the $N_k$ cells of level $k$ per unit time. The
differentiation rate of a single cell is, thus
$\delta_k/N_k$.

In principle five microscopic events can occur with a cell:
(i) symmetric cell division with differentiation,
(ii) asymmetric cell division,
(iii) symmetric cell division without differentiation,
(iv) single cell differentiation,
and (v) cell death (Fig.~\ref{fig2}b).
Our goal is to determine the optimal tissue organization and dynamics that
minimize the number of cell divisions that the cells undergo
until they become terminally differentiated. For this reason cell
death, except for the continuous expenditure of terminally
differentiated cells, is disallowed as it can only increase the number of
divisions.
We note, however, that cell death with a rate proportional
to that of cell divisions would simply result in a proportionally
increased divisional load and, thus, would have no effect on the optimum.

Similarly, we also disregard single cell differentiation, because if
it is rare enough (i.e., its rate is smaller than the asymmetric cell
division rate plus twice the rate of symmetric cell division without
differentiation) then it can be absorbed in 
cell divisions with differentiation; otherwise it would merely delegate the
replication burden down the hierarchy towards the less differentiated
and supposedly less frequently dividing cells, and would be
sub-optimal.

Two of the remaining three microscopic events involve
differentiation. If we denote the fraction of differentiation events
that occur via symmetric cell division at level $k$ by $p_k$, then the
rate of symmetric cell division at level $k$ can be written as
$p_k \delta_k /2$ (the division by $2$
accounts for the two daughter cells produced by a single division),
while the rate of asymmetric cell division is $(1-p_k)\delta_k$.
Symmetric cell division with differentiation leaves an
empty site at level $k$, which will be replenished either
(i) by differentiation from the
level below or (ii) by division on the same level. Assuming the first
case and denoting the
fraction of replenishment events that occur by differentiation from the
level below by $q_k$, the combined rate of the contributing processes
(asymmetric cell division and symmetric cell division with
differentiation from the level below) can be written as $q_k
p_k \delta_k /2$. By definition this is equal
to $\delta_{k-1}$, the differentiation rate from level $k-1$,
leading to the recursion relation
\begin{align}
 \delta_{k-1} &= \delta_k p_k q_k /2
\, .
\end{align}
Alternatively, if replenishment occurs by cell
division on the same level $k$, i.e., as a result of symmetric cell
division without differentiation, the corresponding rate is $(1-q_k)
p_k \delta_k /2$.

To keep track of how cell divisions accumulate along cell lineages during
tissue renewal, we introduce the divisional load $D_k(t)$ for each
level separately defined as the
average number of divisions that cells at level $k$ have undergone by
time $t$ since the stem cell level was created at time zero.

Using the rates of the microscopic events (also shown in Fig.~\ref{fig2}b),
considering that each division increases the accumulated number
of divisions of both daughter cells by one, and taking into
account the divisional loads that the departure of cells
take and the arrival of cells bring, the following mean-field differential
equation system can be formulated for the time evolution of the total
divisional load ($D_k N_k$) of levels $k<n$ of a fully developed tissue:
\begin{align}
 \dot D_k N_k &= - \frac{\delta_k}{2} p_k D_k+ \delta_k (1-p_k)
\nonumber\\
 &+ \frac{\delta_k}{2} p_k \left[ q_k (D_{k-1}+1) + (1-q_k)(D_k+2) \right]
\, .\label{dotD}
\end{align}
Because stem cells cannot be replenished from below we have $q_0=0$. The
terminal level $k=n$ can be included in the system of equations by
specifying $p_n=q_n=1$ and formally defining $\delta_n=2 \delta_{n-1}$.

The above equations are valid when each level $k$ contains the prescribed
number of cells $N_k$ of a fully developed, homeostatic tissue and, therefore,
do not directly describe the initial development of the tissue from the
original stem cells. This shortcoming can, however, be remedied by
introducing virtual cells that at the initial moment ($t=0$) fill up
all $k>0$ levels. As the virtual cells gradually differentiate to
higher levels of the hierarchy, they are replaced by the
descendants of the stem cells. Tissue development is completed when the
non-virtual descendants of the initial stem cell population fill the
terminally differentiated level for the first time,
expelling all virtual cells. Using this approach the initial
development of the tissue is assumed to follow the same dynamics as
the self-renewal of the fully developed tissue.
Even though cell
divisions in a developing tissue might occur at an elevated pace, such
differences in the overall pace of cell divisions
(along with any temporal variation in the tissue dynamics) are irrelevant,
as long as only the relation between the number of cell divisions and
the number of cells generated are concerned.

Using the recursion relation the above differential equation system
simplifies to
\begin{align}
 \dot D_k N_k &= (\delta_k -\delta_{k-1}) - \delta_{k-1} (D_k - D_{k-1})
\, ,
\end{align}
revealing that the average number of cell divisions is independent of both the
fraction of symmetric division $p_k$ in differentiation, and the
fraction of differentiation $q_k$ in replenishment.

From any initial condition $D_k(t)$ converges to the asymptotic solution
\begin{align}
 D_k(t) &= t \frac{\delta_0}{N_0} + D_k^0
\, ,
\label{Dkt}
\end{align}
which shows that the divisional load of the entire tissue grows
linearly according to the differentiation rate of the stem cells
($t\delta_0/N_0$), and the progenitor cells at higher
levels of the hierarchy have an additional load ($D_k^0$) representing
the number of divisions having led to their differentiation. By definition,
the additional load of the stem cells ($D_0^0$) is zero. The convergence
involves a sum of exponentially decaying terms, among which the slowest
one is characterized by the time scale
\begin{align}
 \tau_k^\textrm{tr} &= \sum_{l=1}^k \frac{N_l}{\delta_{l-1}}
\, ,
\label{tau}
\end{align}
which can be interpreted as the transient time needed for the cells at
level $k$ to reach their asymptotic behavior. $\tau_k^\textrm{tr}$ can
also be considered as the transient time required for the initial
development of the tissue up to level $k$. The rationale behind this is
that during development the levels of the hierarchy become populated by the
descendants of the stem cells roughly sequentially, and the initial
population of level $l$ takes about $N_l/\delta_{l-1}$ time after level $l-1$
has become almost fully populated.

Plugging the asymptotic form of
$D_k(t)$ into the system of differential equations and prescribing
$D_0^0=0$, the constants $D_k^0$ can be determined, and expressed as
\begin{align}
 D_k^0 &= \sum_{l=1}^k \frac{\delta_l -\delta_{l-1}}{\delta_{l-1}}
  - \delta_0 \sum_{l=1}^k \frac{N_l}{\delta_{l-1}}
\nonumber\\
 &= \sum_{l=1}^k (\gamma_l - 1) - \frac{\delta_0}{N_0} \tau_k^\textrm{tr}
\, ,
\end{align}
where we have introduced the ratios
\begin{align}
 \gamma_k &= \frac{\delta_k}{\delta_{k-1}} = \frac{2}{p_k q_k} \geq 2
\end{align}
between any two subsequent differentiation rates.
The asymptotic solution then becomes
\begin{align}
 D_k(t) &= \frac{\delta_0}{N_0} (t - \tau_k^\textrm{tr})
  + \sum_{l=1}^k (\gamma_l - 1)
\, .
\label{Dk}
\end{align}
This simple formula, which describes the accumulation of the divisional
load along the levels of a hierarchically organized tissue, is one of
our main results.

\subsection*{Differentiation hierarchies that minimize divisional load}

The number of mutations that a tissue allows for its constituent cells
to accumulate can be best characterized by the expected number of
mutations accumulated along the longest cell lineages. On average, the
longest lineage corresponds to the last terminally differentiated cell
that is produced by the tissue at the end of the lifetime of the
organism. Therefore, as the single most important characteristics of a
hierarchically organized tissue, we define its lifetime divisional
load, $D$, as the divisional load of its last terminally differentiated
cell. If the total number of terminally differentiated cells produced
by the tissue during the natural lifetime of the organism per stem cell
is denoted by $N$, then the lifetime of the organism can be expressed as
$t_\textrm{life} = \tau_{n-1}^\textrm{tr} + N_0 N/\delta_{n-1}$,
where the first term is the development time of the tissue up to level
$n-1$, and the second term is the time necessary to generate all the
$N_0 N$ terminally differentiated cells by level $n-1$ at a rate of
$\delta_{n-1}$. Because the last terminally
differentiated cell is the result of a cell division at level $n-1$,
its expected divisional load, $D$, is the average
divisional load of level $n-1$ increased by $1$:
\begin{align}
 D &=
 D_{n-1} \left( t_\textrm{life} \right) +1
 = N \frac{\delta_0}{\delta_{n-1}}
  + \sum_{l=1}^{n-1} (\gamma_l - 1) + 1
 = N \prod_{l=1}^{n-1} \frac{1}{\gamma_l}
  + \sum_{l=1}^{n-1} (\gamma_l - 1) + 1
\, .
\label{D}
\end{align}
Note that the complicated $\tau_{n-1}^\textrm{tr}$ term drops out of
the formula.
A remarkable property of $D$ is that it
depends only on two structural and two dynamical parameters of the
tissue. The two structural parameters are the total number of the
terminally differentiated cells produced by the tissue per stem cell,
$N$, and the number of the hierarchical levels, $n$. The two dynamical
parameters are the product and sum of the ratios
of the differentiation rates, $\gamma_k$.
The lifetime divisional load neither depends on most of
the microscopic parameters of the cellular processes, nor
on the number of cells at the differentiation levels.

For fixed $N$ and $n$ the ratios $\gamma_k^*$ of the differentiation
rates that minimize the
lifetime divisional load $D$ can be easily determined by setting the
derivatives of $D$ with respect to the ratios $\gamma_k$ to zero,
resulting in
\begin{align}
 \gamma_k^* &= N \prod_{l=1}^{n-1} \frac{1}{\gamma_l^*}
\, .
\end{align}
This expression shows that $\gamma_k^*$ is identical for all
intermediate levels ($0<k<n$) and, therefore, can be denoted by
$\gamma^*$ without a subscript. This uniform ratio can then be expressed as
\begin{align}
 \gamma^* = N^{1/n}
\, ,
\label{gs}
\end{align}
as long as the condition $\gamma^*\geq2$ holds, i.e., when $n\leq\log_2(N)$.
For $n\geq\log_2(N)$, however, the ratio has to take the value of
\begin{align}
 \gamma^* = 2
\, .
\label{gss}
\end{align}
Plugging $\gamma^*$ into Eq.~(\ref{D}) results in
\begin{align}
 D^* &=
 n \left( N^{1/n} - 1 \right) + 2
\label{Ds}
\end{align}
for $n\leq\log_2(N)$ and
\begin{align}
 D^* &=
 N \left( \frac{1}{2} \right)^{n-1} + n
\label{Dss}
\end{align}
for $n\geq\log_2(N)$.
Eq.~(\ref{Ds}) is a monotonically decreasing function of $n$, while
Eq.~(\ref{Dss}) has a minimum at
\begin{align}
 n_\textrm{opt} &=
 \log_2(N) + 1 + \log_2(\ln2) \approx
 \log_2(N) + 0.471
\label{nopt}
\end{align}
levels. This $n_\textrm{opt}$ together with the ratio
\begin{align}
 \gamma^*_\textrm{opt} &= 2
\label{gopt}
\end{align}
represent the optimal tissue-structure in the sense that it minimizes
the lifetime divisional load of a self-renewing tissue, yielding
\begin{align}
 D^*_\textrm{opt} &=
 \log_2(N) + 1 + \log_2(\ln2) + 1/\ln2 \approx
 \log_2(N) + 1.914
\, .
\end{align}
Note that under this optimal condition the divisional rate of the stem
cell level is very low: in a mature tissue (i.e., after the tissue has
developed) the expected number of divisions of a stem cell, which is
equivalent to the expected number of differentiation to level $1$ per
stem cell is only
$(\delta_0/N_0)(N_0 N/\delta_{n-1}) = 1/\ln2 \approx 1.44$.

\subsection*{Implications of the analytical results}

\begin{figure}
\centerline{\includegraphics[width=0.8\columnwidth]{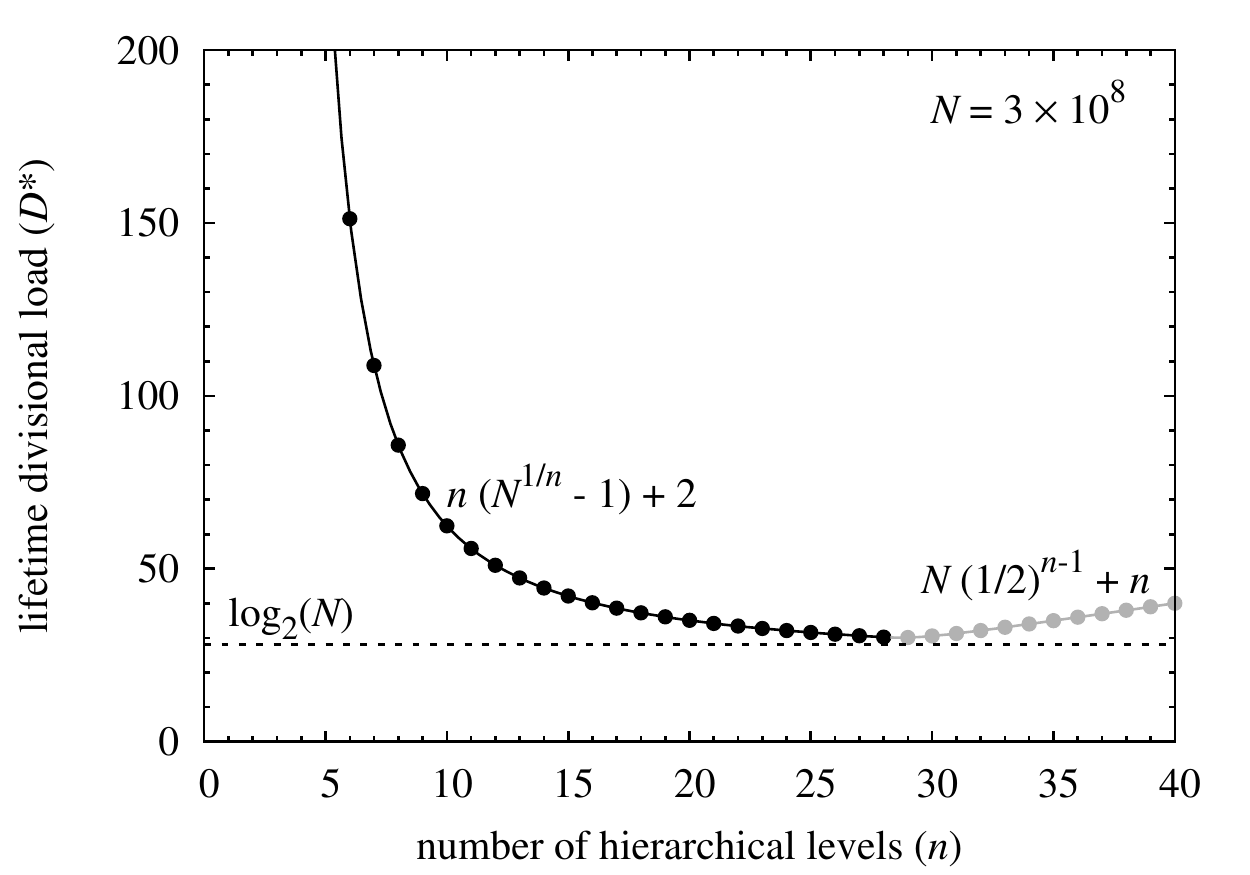}}
\caption{
{\bf The lower limit of the lifetime divisional load as a
function of the number of hierarchical levels.}
The black and gray solid lines (with filled circles at integer values
of $n$) show the lower limit of the lifetime divisional load of a
tissue, $D^*$, as a function of the number of hierarchical levels, $n$,
for $n\leq\log_2(N)$ and $n\geq\log_2(N)$, respectively. The
theoretical minimum, $\log_2(N)$, achievable by a series of divisions
along a perfect binary tree characteristic of non-renewing tissues, is
displayed with a dashed line. Here we have assumed $N=3 \times 10^8$
roughly corresponding to the number of cells shed by
a few square millimeters of human
skin that is sustained by a single stem cell.
}
\label{fig3}
\end{figure}

Remarkably, $D^*_\textrm{opt}$ corresponds to less than two cell divisions in
addition to the theoretical minimum
of $\log_2(N)$, achievable by a series of
divisions along a perfect binary tree characteristic of non-renewing
tissues. In other words, in terms of minimizing the number of necessary cell
divisions along cell lineages, a self-renewing hierarchical tissue
can be almost as effective as a non-renewing one.
Consequently, hierarchical tissue
organization with a sufficient number of hierarchical levels provides
a highly adaptable and practically ideal mechanism not only for ensuring
self-renewability but also
keeping the number of cell divisions near the theoretical absolute
minimum.

An important result of our mathematical analysis is that it provides a
simple and mathematically rigorous formula (Eqs.~\ref{Ds} and
\ref{Dss}, and Fig.~\ref{fig3}) for the lower limit of the lifetime
divisional load of a tissue for a given number of hierarchical levels
and a given number of terminally differentiated cells descending from a
single stem cell. This lower limit can be reached only with a power law
distribution of the differentiation rates (i.e., with a uniform ratio
between the differentiation rates of any two successive differentiation
levels), justifying the assumptions of the models by Werner et al.\
\cite{Werner:2011,Werner:2013}.

In the optimal scenario, where $\gamma_k = \gamma^*_\textrm{opt} = 2$,
the recursion relation imposes $p_n=q_n=1$, thereby, all cell divisions
must be symmetric and involve differentiation. This is a shared feature
with non-renewable differentiation, which is the underlying reason, why
the number of cell divisions of the optimal self-renewing mechanism can
closely approach the theoretical minimum.

As a salient example of self-renewing tissues,
let us consider the human skin.
Clonal patches of skin are of the
order of square millimeters in size \cite{Martincorena:2015}, the top
layer of skin, which is renewed daily,
is composed of approximately a thousand cells per square millimeter
\cite{Hoath:2003}. If we assume that a $10$~mm$^2$ patch is maintained
by a single stem cell for $80$ years, this
corresponds to about $N =3\times 10^8$ cells.
As Fig.~\ref{fig3}
demonstrates, the $D^*$ vs. $n$ curve becomes very flat for large
values of $n$, indicating that in a real tissue the number of
hierarchical levels can be reduced by at least a factor of $2$ from the
optimal value, without significantly compromising the number of
necessary cell divisions along the cell lineages.

It is a question how the total number of terminally differentiated
cells ($N_0 N$) produced by the tissue during the natural lifetime of
the organism can be best partitioned into the number of tissue-specific
stem cells ($N_0$) and the number of terminally differentiated cells per
stem cell ($N$). The initial generation of the stem cells along a
binary tree requires $\log_2(N_0)$ divisions. The production of the
terminally differentiated cells in a near-optimal hierarchy
requires about $\log_2(N)$ divisions. Their sum, which is about
$\log_2(N_0 N)$, depends only on the total number of terminally
differentiated cells, irrespective of the number of stem cells.
This means, that the minimization of the divisional load poses no constraint
on the number of stem cells. However, since both maintaining a larger
number of differentiation levels and keeping the differentiation
hierarchy closer to optimum involve more complicated regulation,
we suspect that a relatively large stem cell pool is beneficial,
especially as a larger stem cell population can also be expected to
be more robust against stochastic extinction, population oscillation,
and injury.

\section*{Discussion}

In general, how closely the hierarchical organization of different
tissues in different organisms approaches the optimum described above
depends on (i) the strength of natural selection against unwanted somatic
evolution, which is expected to be much stronger in larger and longer
lived animals; and (ii) intrinsic physiological constraints on the
complexity of tissue organization and potential lower limits on stem
cell division rate. Neither the strength of selection nor the
physiological constraints on tissue organization are known at
present. However, in the case of the germ line mutation rate,
which is proportional to the number of cell divisions in lineages
leading to the gametes, current evidence indicates that physiological
constraints are not limiting \cite{Lynch:2012}.
Across species,
differences in effective population size, which is in general negatively
correlated with body size and longevity \cite{Nabholz:2013}, indicate
the effectiveness of selection relative to drift. As a result,
differences in effective population size between species determine the
effectiveness of selection in spreading of favorable
mutations and eliminating deleterious ones and, as such,
can be used as indicator of the efficiency of selection
\cite{Kimura:1983,Charlesworth:2009}.
This implies that, in contrast to somatic tissues,
we expect germ line differentiation hierarchies to be more optimal
for smaller animals with shorter life spans as a result of their
increased effective population sizes.
For species for which information is available, the number of
levels across species indeed follows an
increasing trend as a function of the effective population size,
ranging from $n=5$ in humans with relatively small
effective population size of approximately $10^4$
and correspondingly less efficient selection,
$n=8$ in macaque with intermediate effective population size of the
order of $10^5$, and
$n=10$ in mice with the largest effective population
size of approximately $5\times10^5$ \cite{Lynch:2010,Ramm:2014}.

A qualitative examination of
Fig.~\ref{fig3} suggests that a similar number of levels, of the order
of $n\approx10$ may be present in most somatic tissues, because the
$D^*$ vs. $n$ curve becomes progressively flatter after it reaches
around twice the optimal value of $D^*$ at $n\gtrsim10$, and the reduction
in the divisional load becomes smaller and smaller as additional
levels are added to the hierarchy
and other factors are expected to limit further increase in $n$.
Alternatively, if we consider for example the human hematopoietic
system, where approximately $10^4$ hematopoietic stem cells (HSCs)
produce a daily supply of
$\sim3.5\times10^{11}$ blood cells, we can calculate that over
$80$ years each stem cell produces a total of $N\approx10^{12}$
terminally differentiated cells. For this larger value of $N$ the
$D^*$ vs. $n$ curve reaches twice the optimal value of
$D^*$ at $n\gtrsim15$ after which, similarly to Fig.~\ref{fig3}, it
becomes progressively flatter and the reduction in divisional load
diminishes as additional levels are added. This rough estimate of
$n\gtrsim15$ levels is consistent with explicit mathematical models of
human hematopoiesis that predict between $17$ and $31$ levels
\cite{Dingli:2007}.
Active or short term HSCs (ST-HSCs) are estimated to differentiate about
once a year, whereas a quiescent population of HSCs that provides cells
to the active population is expected to be characterized by an even
lower rate of differentiation. This is in good agreement with our
prediction about the existence of a heterogeneous stem cell pool,
a fraction of which consists of quiescent cells that only undergo a
very limited number of cell cycles during the lifetime of the organism.
Indeed, recently Busch et al.\ found that adult
hematopoiesis in mice is largely sustained by
previously designated ST-HSCs that nearly fully
self-renew, and receive rare but polyclonal HSC input
\cite{Busch:2015}. Mouse HSCs were found to differentiate into ST-HSCs only about
three times per year.

For most somatic tissues the differentiation hierarchies that underpin
the development of most cellular compartments remain inadequately
resolved, the identity of stem and progenitor cells remains uncertain,
and quantitative information on their proliferation rates is limited
\cite{Sutherland:2015}. However, synthesis of available information on
tissue organization by Tomasetti and Vogelstein \cite{Tomasetti:2015},
as detailed above, suggests that larger tissues with rapid turnover
(e.g., colon and blood) are relatively less prone to develop cancer.
This phenomenon, as noted in the introduction, can be interpreted as Peto's paradox across tissues
with the implication that larger tissues with rapid turnover rates have
hierarchies with more levels and stem cells that divide at a slower
pace. Accumulating evidence from lineage-tracing experiments
\cite{Blanpain:2013}
is also consistent with a relatively large number of hierarchical levels.
Populations of stem cells in blood, skin, and the
colon have begun to be resolved as combinations of cells that are
long-lived yet constantly cycling,
and emerging
evidence indicates that both quiescent and active cell subpopulations
may coexist in several tissues, in separate yet adjoining locations
\cite{Li:2010}.
Lineage-tracing techniques \cite{Blanpain:2013} are rapidly
developing, and may be used for directly testing the predictions of our
mathematical model about the highly inhomogeneous distributions of
the differentiation rates in the near future. In the context of estimates of
the number of stem cells in different tissues that underlie
Tomasetti and Vogelstein's results, the potential existence of such
unresolved hierarchical levels suggests the possibility that the number
of levels of the hierarchy are systematically underestimated and,
correspondingly, that the number of stem cells at the base of these
hierarchies are systematically overestimated.

Independent of the details of the hierarchy the dynamics of how
divisional load accumulates in time is described by two phases:
(i) a transient development phase during which each level of the
hierarchy is filled up and
(ii) a stationary phase during which homeostasis is maintained in
mature tissue.
The dynamic details and the divisional load incurred during the initial
development phase depend on the details of the hierarchy (cf.\
Eqs.~(\ref{Dk}) and (\ref{tau})). In contrast, in the stationary
phase, further accumulation of the mutational load is determined solely
by $\delta_0/N_0$ the rate at which tissue-specific stem cells differentiate at
the bottommost level of the hierarchy. Such biphasic behavior has been
observed in the accumulation of mutations both in somatic
\cite{Rozhok:2015} and germ line cells
\cite{Kong:2012,Gao:2016,Rahbari:2016}. In both cases a substantial
number of mutations were found to occur relatively rapidly during
development followed by a slower linear accumulation of mutation
thereafter. General theoretical arguments imply that
the contribution of the mutational load incurred during development to
cancer risk is substantial \cite{Frank:2003}, but this has been suggested to be in
conflict with the fact that the majority of cancers develop late in
life \cite{Rozhok:2015,Rozhok:2016}. Resolving this question and more
generally understanding the development of cancer in self-renewing
tissues will require modeling the evolutionary dynamics of how the
hierarchical organization of healthy tissues breaks down.

Spontaneously occurring mutations accumulate in somatic
cells throughout a person's lifetime, but the majority of these
mutations do not have a noticeable effect. A small minority,
however, can alter key cellular functions and a fraction of
these confer a selective advantage to the cell, leading to preferential
growth or survival of a clone \cite{Martincorena:2015rev}. Hierarchical tissue
organization can limit somatic evolution at both these levels: (i) at the
level of mutations, as we
demonstrated above, it can dramatically reduce the number of cell
divisions required and correspondingly the mutational load incurred
during tissue homeostasis; and (ii) at the level of selection acting
on mutations with non-neutral phenotypic effects, as demonstrated by
Nowak et al.\ \cite{Nowak:2003} and later by Pepper et al. \cite{Pepper:2007},
tissues organized into serial differentiation
experience lower rates of such detrimental cell-level phenotypic evolution.
Extending the seminal results of Nowak et al.\ and Pepper et al., we
propose that in addition to limiting somatic evolution at the
phenotypic level, hierarchies are also how the tissues of multicellular
organisms
keep the accumulation of mutations in check, and that tissue-specific
stem cells may in general correspond to a diverse set of slower
dividing cell types.

In summary, we have considered a generic model of hierarchically
organized self-renewing tissue, in the context of which we have derived
universal properties of the divisional load during tissue homeostasis.
In particular, our results provide a lower bound for the
lifetime divisional load of a tissue as a function of the number of its
hierarchical levels.
Our simple analytical description provides a
quantitative understanding of how hierarchical tissue organization can
limit unwanted somatic evolution, including cancer development.
Surprisingly, we
find that the theoretical minimum number of cell divisions can be
closely approached (cf.\ Fig.~\ref{fig3}, where the theoretical minimum
corresponds to the dashed horizontal line), demonstrating that hierarchical
tissue organization
provides a robust and nearly ideal mechanism to limit the divisional
load of tissues and, as a result, minimize somatic evolution.

\begin{acknowledgments}

This work was supported by the Hungarian Science Foundation (grant K101436).
%
%
The authors would like to acknowledge the comments of anonymous
reviewers on a previous version of the manuscript, as well as discussion
with and comments from Bastien Boussau, M\'arton Demeter, M\'ate Kiss, and
D\'aniel Grajzel.
\end{acknowledgments}

\section*{Data availability}
No data was generated as part of this study.

\section*{Conflict of Interest}
The authors declare no conflict of interest.

\section*{Author contributions}

I.D. and Sz.G. designed the study, carried out research, and wrote the paper.

\bibliographystyle{plainurl}
\bibliographystyle{unsrt}

\end{document}